\documentclass[aps,prd,twocolumn,superscriptaddress]{revtex4}

\usepackage{graphicx}% Include figure files
\usepackage[hypertex]{hyperref}
\def\eg{{\it e.g.},~}
\def\beq{\begin{equation}}
\def\eeq{\end{equation}}
\def\bea{\begin{eqnarray}}
\def\eea{\end{eqnarray}}

\def\msun{M_{\odot}}
\usepackage{color}

\begin{document}

\title{Reviving Gravity's Aether in Einstein's Universe}

\author{Niayesh Afshordi}\email{nafshordi@perimeterinstitute.ca}
\affiliation{Perimeter Institute
for Theoretical Physics, 31 Caroline St. N., Waterloo, ON, N2L 2Y5,Canada}
\affiliation{Department of Physics and Astronomy, University of Waterloo, 200 University Avenue West, Waterloo, ON, N2L 3G1, Canada }

%\date{\today}
%\preprint{astro-ph/yymmnnn}

\begin{abstract}
Einstein's theory of general relativity describes gravity as the interaction of particles with space-time geometry, as opposed to interacting with a physical fluid, as in the old gravitational aether theories. Moreover, any theoretical physicist would tell you that, despite its counter-intuitive structure, general relativity is one of the simplest, most beautiful, and successful theories in physics, that has withstood a diverse battery of precision tests over the past century. So, is there any motivation to relax its fundamental principle, and re-introduce a gravitational aether? Here, I give a short and non-technical account of why quantum gravity and cosmological constant problems provide this motivation.
\end{abstract}

\maketitle

%\section{Introduction}

Ask any good student of freshman physics and, happily quoting their textbooks, they will tell you that gravity is the weakest force of nature.
After all, when you lift a pen, the electromagnetic dipoles of the molecules in your hand can easily counteract the gravitational pull from the entire planet Earth. It may thus come as a surprise that throughout history, understanding gravity has been one of the strongest drivers of breakthroughs in theoretical physics, and yet it still remains its deepest mystery.

After Newton's discovery of universal laws of gravity and mechanics, physicists and philosophers often wondered how gravitational forces could act over large distances, while other forces of nature only act in extreme proximity. In fact, this was one of Einstein's philosophical motivations to introduce metric, or space-time geometry, as a medium that mediates gravitational forces, as ``action at a distance'' cannot be physical. But we are jumping ahead of ourselves!

 Long before Einstein's celebrated invention of General Relativity, over the course of the 16th to 19th centuries, many mechanical models of gravity were put forth and then discarded. In these theories, an invisible medium, called ``the gravitational aether'',  mediated the particles, vortices, streams, or waves that exchanged gravitational force between massive bodies \footnote{Mechanical explanations of gravitation, {\it Wikipedia.org}, \href{http://en.wikipedia.org/wiki/Mechanical_explanations_of_gravitation}{ http://en.wikipedia.org/wiki/Mechanical\_ explanations\_ of\_ gravitation}.}. For example, in 1853, Riemann proposed that gravitational aether was an incompressible fluid which sinks toward massive objects where it is absorbed, at a rate proportional to their mass. He speculated that the absorbed aether is then emitted into another spatial dimension \cite{Riemann}.

%\footnote{ Riemann, B. (1876), "Neue mathematische Prinzipien der Naturphilosophie", \href{http://quod.lib.umich.edu/cgi/t/text/pageviewer-idx?c=umhistmath&cc=umhistmath&idno=abw1043.0001.001&frm=frameset&view=image&seq=6}{Bernhard Riemanns Werke und gesammelter Nachlass} (Leipzig): 528-538}.

The most famous refutation of aether theories (even though it did not directly concern the gravitational aether) came from the
Michelson-Morley experiment \cite{Michelson},
%\footnote{Michelson, Albert Abraham \& Morley, Edward Williams (1887), "On the Relative Motion of the Earth and the Luminiferous Ether", American Journal of Science 34: 333345},
which showed that the speed of light is constant, and independent of reference frame, as opposed to being only constant and isotropic in the aether's frame of reference. Indeed, the absence of a preferred reference frame, otherwise known as the principle of relativity, was the key assumption in the development of special, and then general relativity.

\section{Reviving the incompressible gravitational aether}
Einstein's theory of general relativity describes gravity as the interaction of particles with space-time geometry, as opposed to interacting with a physical fluid, as in the old gravitational aether theories. Moreover, any theoretical physicist would tell you that, despite its counter-intuitive structure, general relativity is one of the simplest, most beautiful, and successful theories in physics, that has withstood a diverse battery of precision tests over the past century. So, is there any motivation to relax its fundamental principle, and re-introduce a gravitational aether?

Let us consider an interesting analogy with Newtonian gravity. A hypothetical 19th century philosopher, Dr. John Smith, proposes that the laws of gravity are set by three fundamental principles:
\vspace{0.2cm}

\noindent 1-Bound orbits in the two-body problem must be closed.\\
2-There exist unbound orbits in the two-body problem.\\
3-Gravitational forces obey linear superposition.
\vspace{0.2cm}

These principles uniquely fix the formulation of Newtonian gravity and celestial mechanics. However, we now know that Principle (1), which fixes the inverse square law \footnote{The only closed orbits in a spherical force field are for inverse square, and linear forces. However, the latter do not have unbound orbits, thus violating Principle (2).}, is based on an accidental symmetry between radial and angular frequencies. General relativity violates this symmetry, which is the origin of Mercury's anomalous perihelion precession. Nevertheless, Dr. Smith would have ruled out Einstein's general relativity, as it did not respect his fundamental principles of gravitational theory, as stated above.

The lesson from this story is that the underlying principles or symmetries of an effective theory might be accidental or emergent symmetries of a more fundamental theory. As powerful as the principle of relativity might have been in the development of Einstein's theory of gravity, it might need to be broken/re-examined, \eg by having a preferred reference frame, or a gravitational aether, in a more complete theory of gravity.

But is there any reason to think that general relativity is not the fundamental theory of gravity?

The main motivation for this comes from quantum mechanics, the other hugely successful physical theory of the 20th century:
both general relativity and quantum mechanics have been incredibly successful in describing macroscopic and microscopic phenomena respectively. However, any attempt to apply the rules of quantum mechanics to general relativity seems to lead to divergences that impair the predictive power of the theory. The effective theory of gravity breaks down when the macroscopic and microscopic worlds meet and a huge amount of energy is packed into small scales, i.e., energy densities exceeding the Planck density of $10^{114}$ Joules (or $10^{97}$ kilograms) per cubic meter. Although it is hard to achieve such densities in laboratories, Penrose and Hawking \cite{1970RSPSA.314..529H} showed that singularities with infinite densities are inevitable in the future and past of general relativistic dynamics. While they may not be immediately accessible to us, they should be prevalent in the universe, residing at the centers of millions of astrophysical black holes in our galaxy, and possibly present at the first moment of the cosmological big bang. It is generally believed that a fundamental theory of quantum gravity should give a self-consistent description of physics close to these singularities (and thus avoid their formation). General relativity plus  quantum mechanics does not.

Most physicists agree on the status of the problem at this level. However,  they diverge on their approaches from this point on. One approach is the interesting possibility of relaxing the requirement of no preferred reference frame (or Lorentz invariance). While the geometric nature of gravity is ubiquitous, there might still exist a physical gravitational aether, which only interacts with geometry (or matter) at very high energies.

 Recently Petr Ho${\rm \check{r}}$ava generated a lot of excitement by suggesting that if the speed of propagation of gravitons increases with energy as $E^{2/3}$ at very high energies, then the theory of gravity might have a well-defined quantization \cite{Horava:2009uw}. This of course introduces a preferred frame in which the energy $E$ is measured.

While breaking Lorentz invariance may sound heretical to many physicists, it comes easily to cosmologists. After all, even though our laws don't seem to have a preferred frame of reference, the universe hasn't had much trouble in picking one. For example, a relativistic electron in the universe will eventually come to stop in the rest frame of the cosmic microwave background (CMB), where the CMB dipole vanishes. That is why analogues of the invisible aether, such as dark matter, dark energy, and the inflaton exist and play crucial roles in the standard model of cosmology.

While it is typical to {\it spontaneously} break Lorentz symmetry on cosmological  scales, normal matter on very small scales/high energies decouples from this cosmological frame. Nevertheless, it is easy to find theories that do not behave this way, and yet are consistent, at least up to some high energy cut-off. For scalar field theories, this can be done through covariant actions that are not quadratic in field gradients. An extreme example of this is the ``cuscuton action'' \cite{Afshordi:2006ad,Afshordi:2007yx}, defined as:
\beq
S = \int d^4x \sqrt{-g} \left[\mu^2 \sqrt{\partial_\mu \varphi \partial^\mu \varphi } - V(\varphi)\right],
\eeq
which represents an incompressible fluid, implying that perturbations around any uniform density background are non-dynamical. It is interesting to note that Ho${\rm \check{r}}$ava's gravity theory reduces to general relativity minimally coupled to an incompressible cuscuton fluid at low energies \cite{Afshordi:2009tt}.

At this point, it is interesting to recall Riemann's idea of an ``incompressible gravitational aether'', and to entertain the possibility that after 156 years, it might turn out to be an actual ingredient of a quantum theory of gravity. There are, after all, no new ideas under the sun!

\section{gravitational aether and the cosmological constant problem}

Although one may decide to ignore the problem of quantizing gravity for low energy and large scale observations, there is one aspect of quantum mechanics that is disastrous for any gravitational observable: the quantum vacuum of the standard model of particle physics has a density of roughly $10^{33}$ kilograms per cubic meter! One does not need precision observations to conclude that this is not realistic, as human bodies, let alone stars and planets would be torn apart by extreme gravitational tidal forces. Incidentally, there are cosmological precision measurements of the vacuum density, which put it at \cite{Dunkley:2008ie}:
\beq
\rho_{\rm vac} = (7.1 \pm 0.9) \times 10^{-27}~ {\rm kg/m^3},\label{rhovac}
\eeq
i.e. some 60 orders of magnitude smaller than the standard model prediction! Of course, there could be other unknown contributions to the vacuum density, but why should they so precisely (but not completely) cancel the known contributions? This is known as the cosmological constant problem.

One way to avoid the problem is to couple gravity to the traceless part of the energy-momentum tensor, effectively decoupling the vacuum energy from gravity:
\beq
(8\pi G')^{-1}G_{\mu\nu}[g_{\mu\nu}]= T_{\mu\nu} -\frac{1}{4}T^{\alpha}_{\alpha} g_{\mu\nu} + T'_{\mu\nu}. \label{gmod}
\eeq
Eq. (\ref{gmod}) is a modification of the celebrated Einstein equation, which couples the space-time curvature, represented by the Einstein tensor, $G_{\mu\nu}$ on the left, to the matter energy-momentum tensor $T_{\mu\nu}$ on the right. However, the last two terms on the right are new: the second term subtracts the trace of $T_{\mu\nu}$, which effectively decouples the vacuum from gravity. The last term is there to ensure energy-momentum conservation $T^{\nu}_{\mu;\nu} =0$, as Bianchi identity enforces zero divergence for the Einstein tensor $G^{\nu}_{\mu;\nu} =0$. Therefore, we require
\beq
T'^{\nu}_{\mu;\nu} = \frac{1}{4} T^{\nu}_{\nu,\mu}.
\eeq

$T'_{\mu\nu}$ is a new component of gravitational dynamics, which we can think of as a modern-day version of the gravitational aether \cite{Afshordi:2008xu}. Moreover, through the above argument,  it is an inevitable component of a complete theory of gravity if we decide to decouple the quantum vacuum energy from geometry.

Of course, one needs to know more about the properties of aether in order to make predictions in this theory. By now, it may not come to the reader as a surprise that we shall assume aether to be incompressible, or more specifically, to have zero density, but non-vanishing pressure. The main motivation, apart from its historical appeal and appearance in quantum gravity theories, is that an incompressible fluid does not introduce new dynamical degrees of freedom, which are severely constrained by precision tests of gravity.

What is surprising about this theory is how similar its predictions are to those of general relativity. In fact, the two are only significantly different in objects with relativistic pressure (such as neutron stars, or the early universe) or large vorticity \cite{Afshordi:2008xu}. The main effect of the new terms on the right hand side of Eq. (\ref{gmod}) is to create an effective Newton's constant which depends on the equation of state of matter, $w_{\rm matt.}=p_{\rm matt.}/\rho_{\rm matt.}$, the ratio of pressure to density:
\beq
G_{\rm eff} \simeq (1+w_{\rm matt.}) G_{N}.
\eeq
While this change is negligible in most astrophysical situations, it significantly changes the dynamics of the early universe, as the gravitation due to radiation is enhanced by a factor of $4/3$. To a good approximation, this effect can be captured in the standard cosmological model by increasing the number of neutrinos from 3 to 5.5, while keeping the gravitational constant fixed. Surprisingly, this is exactly what is found in analysis of the Lyman-$\alpha$ forest in quasar spectra ($N_{\nu} = 5.3 \pm 1.1$), even though it is marginally inconsistent with observational constraints on big bang nucleosyntheis \cite{Seljak:2006bg}. Future cosmological observations will be able to rule out or confirm this prediction conclusively.

There is one final question that might be lingering in the reader's mind. If gravity is completely decoupled from the vacuum energy, how could we measure the vacuum density as having the value in Eq. (\ref{rhovac})? This measurement is based on the observation that the cosmic expansion appears to have started accelerating about 6 billion years ago. The easiest way to explain this is a uniform vacuum energy density which dominates today's cosmic energy density, and is amazingly consistent with almost all the cosmological observations.

\begin{figure}
\includegraphics[width=\linewidth]{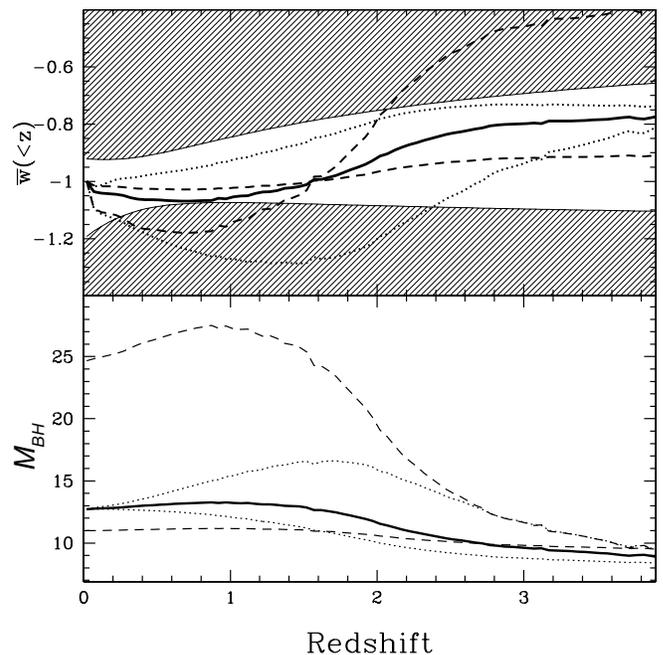}
 \caption
{From \cite{PrescodWeinstein:2009mp}: {\bf Top panel:} The prediction of
   different astrophysical black hole formation scenarios (see below) for the effective dark energy equation of state $\bar{w}(<z)$, given that aether pressure scales as inverse cube of the mean black hole mass, $M^{-3}_{BH}$. This can be compared to constraints from cosmology.  The unshaded area shows the region currently allowed at 68\% confidence level for this
parameter, as measured from cosmological observations \cite{Komatsu:2008hk}.
{\bf Bottom panel:} The mass-weighted geometric mean of black hole
masses, $M_{BH}$, in units of $\msun$ as a function of redshift.   Our fiducial model (solid, black line) assumes our best
estimates of the mass distribution evolution of the black hole
mass distribution. Dashed lines indicate the range of uncertainty
expected due to the unknown relative contribution of
supermassive and stellar-mass black holes, while
the dotted lines represent the uncertainty in the shape of the
star formation density evolution. These correspond to the same models used in the top panel.
}\label{fig-obsmstar}
\end{figure}

It turns out that a similar phenomenon happens as black holes form in the gravitational aether scenario. While in general relativity, formation of singularities is shielded from the outside  world by event horizons, the incompressible gravitational aether with an {\it infinite speed of sound} is not bound by the horizons. Therefore, the onset of the quantum gravity regime close to the singularity might affect aether pressure outside the black hole. In \cite{PrescodWeinstein:2009mp,Afshordi:2010eq}, it was shown that an incompressible gravitational aether ties the geometry close to the black hole horizon to cosmological scales. Assuming Planck scale physics close to the horizon, one can show that the pressure of aether at infinity roughly scales as $M_{BH}^{-3}$, and is comparable to today's vacuum pressure for $M_{BH} = 10-100 M_{\odot}$. Incidentally, this is the typical mass range for stellar black holes in our universe. Therefore, the gravitational aether scenario could potentially explain today's acceleration of cosmic expansion, {\it without any fine-tuning}, by virtue of a quantum gravitational effect close to the horizon of stellar black holes. Furthermore, Fig. \ref{fig-obsmstar} shows that this model makes concrete predictions for the evolution of cosmic acceleration over time, that appear to match well with current observations. Future observational probes of cosmic acceleration and galaxy formation will be able to definitively rule out or confirm this proposed connection between dark energy and astrophysical black holes over the next decade.

\section{Conclusions}

Unifying general relativity and quantum mechanics, the two great physical theories of the twentieth century, has fascinated and puzzled theoretical physicists for many decades. As bizarre as it may sound, recycling discarded ideas of the 19th century might provide a way forward!

While gravitational aether is far from the only possibility for solving the problems of quantum gravity, the theoretical arguments and motivations for  its reincarnation are simple and sound, and the coincidence of its predictions with cosmological observations is very suggestive. Many questions still remain, and need to be answered in order to have a viable physical theory on par with general relativity:
Is there an action for this theory with a well-defined quantization? Can a UV completion of the theory resolve the structure of black hole horizons?
 What does black hole formation look like in this theory? Will there be smoking guns in the future precision tests of gravity? Is aether consistent with all cosmological observations? What about the anomalies such as those in the integrated Sachs-Wolfe \cite{Ho:2008bz} and large-angle CMB anisotropies \cite{Copi:2008hw}?

Looking forward, one expects the revival of gravitational aether to lead to many new possibilities in our theoretical understanding of quantum gravity and quantum cosmology, as well as the phenomenology of astrophysical and cosmological observations. The resolution of last century's mysteries may not be too far off after all.

\vspace{1cm}
Research at Perimeter Institute is
supported by the Government of Canada through Industry Canada and by the
Province of Ontario through the Ministry of Research \& Innovation.

%\bibliographystyle{utcaps_na2_ads}
%\bibliographystyle{utcaps_na2}
%\bibliography{revival_aether_2.bbl}

\begin{thebibliography}{10}

\bibitem{Riemann}
B.~Riemann, {\em ``Neue mathematische Prinzipien der Naturphilosophie'',
  Bernhard Riemanns Werke und gesammelter Nachlass} (1876)  528--538.

\bibitem{Michelson}
A.~A. Michelson and E.~W. Morley, {\em {American Journal of Science}} {\bf
  34} (1887)  333--345.

\bibitem{1970RSPSA.314..529H}
S.~W. {Hawking} and R.~{Penrose}, {\em Royal Society of London Proceedings
  Series A} {\bf 314} (Jan., 1970)  529--548.

\bibitem{Horava:2009uw}
P.~Horava, \href{http://dx.doi.org/10.1103/PhysRevD.79.084008}{{\em Phys.
  Rev.} {\bf D79} (2009)  084008},
\href{http://arxiv.org/abs/0901.3775}{{\tt arXiv:0901.3775 [hep-th]}}.
%%CITATION = 0901.3775;%%.

\bibitem{Afshordi:2006ad}
N.~Afshordi, D.~J.~H. Chung, and G.~Geshnizjani,
  \href{http://dx.doi.org/10.1103/PhysRevD.75.083513}{{\em Phys. Rev.} {\bf
  D75} (2007)  083513},
\href{http://arxiv.org/abs/hep-th/0609150}{{\tt arXiv:hep-th/0609150}}.
%%CITATION = HEP-TH/0609150;%%.

\bibitem{Afshordi:2007yx}
N.~Afshordi, D.~J.~H. Chung, M.~Doran, and G.~Geshnizjani,
  \href{http://dx.doi.org/10.1103/PhysRevD.75.123509}{{\em Phys. Rev.} {\bf
  D75} (2007)  123509},
\href{http://arxiv.org/abs/astro-ph/0702002}{{\tt arXiv:astro-ph/0702002}}.
%%CITATION = ASTRO-PH/0702002;%%.

\bibitem{Afshordi:2009tt}
N.~Afshordi, \href{http://dx.doi.org/10.1103/PhysRevD.80.081502}{{\em Phys.
  Rev.} {\bf D80} (2009)  081502},
\href{http://arxiv.org/abs/0907.5201}{{\tt arXiv:0907.5201 [hep-th]}}.
%%CITATION = 0907.5201;%%.

\bibitem{Dunkley:2008ie}
J.~Dunkley {\em et al.},
  \href{http://dx.doi.org/10.1088/0067-0049/180/2/306}{{\em Astrophys. J.
  Suppl.} {\bf 180} (2009)  306--329},
\href{http://arxiv.org/abs/0803.0586}{{\tt arXiv:0803.0586 [astro-ph]}}.
%%CITATION = 0803.0586;%%.

\bibitem{Afshordi:2008xu}
N.~Afshordi,
\href{http://arxiv.org/abs/0807.2639}{{\tt arXiv:0807.2639 [astro-ph]}}.
%%CITATION = 0807.2639;%%.

\bibitem{Seljak:2006bg}
U.~Seljak, A.~Slosar, and P.~McDonald, {\em JCAP} {\bf 0610} (2006)  014,
\href{http://arxiv.org/abs/astro-ph/0604335}{{\tt arXiv:astro-ph/0604335}}.
%%CITATION = ASTRO-PH/0604335;%%.

\bibitem{PrescodWeinstein:2009mp}
C.~Prescod-Weinstein, N.~Afshordi, and M.~L. Balogh,
 \href{http://dx.doi.org/10.1103/PhysRevD.80.043513}{{\em Phys. Rev.} {\bf
  D80} (2009)  043513},
\href{http://arxiv.org/abs/0905.3551}{{\tt arXiv:0905.3551 [astro-ph.CO]}}.
%%CITATION = 0905.3551;%%.

\bibitem{Afshordi:2010eq}
  N.~Afshordi,
  \href{http://arxiv.org/abs/1003.4811}{{\tt arXiv:1003.4811 [hep-th]}}.


\bibitem{Komatsu:2008hk}
E.~Komatsu {\em et al.},
  \href{http://dx.doi.org/10.1088/0067-0049/180/2/330}{{\em Astrophys. J.
  Suppl.} {\bf 180} (2009)  330--376},
\href{http://arxiv.org/abs/0803.0547}{{\tt arXiv:0803.0547 [astro-ph]}}.
%%CITATION = 0803.0547;%%.

\bibitem{Ho:2008bz}
S.~Ho, C.~Hirata, N.~Padmanabhan, U.~Seljak, and N.~Bahcall,
  \href{http://dx.doi.org/10.1103/PhysRevD.78.043519}{{\em Phys. Rev.} {\bf
  D78} (2008)  043519},
\href{http://arxiv.org/abs/0801.0642}{{\tt arXiv:0801.0642 [astro-ph]}}.
%%CITATION = 0801.0642;%%.

\bibitem{Copi:2008hw}
C.~J. Copi, D.~Huterer, D.~J. Schwarz, and G.~D. Starkman,
  \href{http://dx.doi.org/10.1111/j.1365-2966.2009.15270.x}{{\em Mon. Not. Roy.
  Astron. Soc.} {\bf 399} (2009)  295--303},
\href{http://arxiv.org/abs/0808.3767}{{\tt arXiv:0808.3767 [astro-ph]}}.
%%CITATION = 0808.3767;%%.

\end{thebibliography}

\providecommand{\href}[2]{#2}\begingroup\raggedright\endgroup

\end{document}